\documentclass{paperClass}
\usepackage{multirow}
\usepackage{array}
\newcolumntype{R}[1]{>{\raggedleft\arraybackslash}p{#1}}
\begin{document}
\newcommand{\berlin}{Institut f\"{u}r Physik, Humboldt-Universit\"{a}t zu Berlin, Newtonstr.~~15, 12489 Berlin, Germany}
\newcommand{\bonn}{Helmholtz-Institut f\"{u}r Strahlen- und Kernphysik, Rheinische Friedrich-Wilhelms-Universit\"{a}t Bonn, Nussallee 14-16, 53115 Bonn, Germany}
\newcommand{\cmu}{Department of Physics, Carnegie Mellon University, Pittsburgh, PA 15213, USA}
\newcommand{\cwru}{Department of Physics, Case Western Reserve University, Cleveland, OH 44106, USA}
\newcommand{\etp}{Institute of Experimental Particle Physics~(ETP), Karlsruhe Institute of Technology~(KIT), Wolfgang-Gaede-Str.~1, 76131 Karlsruhe, Germany}
\newcommand{\fulda}{University of Applied Sciences~(HFD)~Fulda, Leipziger Str.~123, 36037 Fulda, Germany}
%
\newcommand{\iap}{Institute for Astroparticle Physics~(IAP), Karlsruhe Institute of Technology~(KIT), Hermann-von-Helmholtz-Platz 1, 76344 Eggenstein-Leopoldshafen, Germany}
\newcommand{\ipe}{Institute for Data Processing and Electronics~(IPE), Karlsruhe Institute of Technology~(KIT), Hermann-von-Helmholtz-Platz 1, 76344 Eggenstein-Leopoldshafen, Germany}
\newcommand{\itep}{Institute for Technical Physics~(ITEP), Karlsruhe Institute of Technology~(KIT), Hermann-von-Helmholtz-Platz 1, 76344 Eggenstein-Leopoldshafen, Germany}
\newcommand{\ppq}{Project, Process, and Quality Management~(PPQ), Karlsruhe Institute of Technology~(KIT), Hermann-von-Helmholtz-Platz 1, 76344 Eggenstein-Leopoldshafen, Germany    }
%
%
\newcommand{\inr}{Institute for Nuclear Research of Russian Academy of Sciences, 60th October Anniversary Prospect 7a, 117312 Moscow, Russia}
\newcommand{\lbnl}{Institute for Nuclear and Particle Astrophysics and Nuclear Science Division, Lawrence Berkeley National Laboratory, Berkeley, CA 94720, USA}
\newcommand{\madrid}{Departamento de Qu\'{i}mica F\'{i}sica Aplicada, Universidad Autonoma de Madrid, Campus de Cantoblanco, 28049 Madrid, Spain}
\newcommand{\mainz}{Institut f\"{u}r Physik, Johannes-Gutenberg-Universit\"{a}t Mainz, 55099 Mainz, Germany}
\newcommand{\mpp}{Max-Planck-Institut f\"{u}r Physik, F\"{o}hringer Ring 6, 80805 M\"{u}nchen, Germany}
\newcommand{\massit}{Laboratory for Nuclear Science, Massachusetts Institute of Technology, 77 Massachusetts Ave, Cambridge, MA 02139, USA}
\newcommand{\mpik}{Max-Planck-Institut f\"{u}r Kernphysik, Saupfercheckweg 1, 69117 Heidelberg, Germany}
\newcommand{\muenster}{Institute for Nuclear Physics, University of M\"{u}nster, Wilhelm-Klemm-Str.~9, 48149 M\"{u}nster, Germany}
\newcommand{\npi}{Nuclear Physics Institute,  Czech Academy of Sciences, 25068 \v{R}e\v{z}, Czech Republic}
\newcommand{\unc}{Department of Physics and Astronomy, University of North Carolina, Chapel Hill, NC 27599, USA}
\newcommand{\washington}{Center for Experimental Nuclear Physics and Astrophysics, and Dept.~of Physics, University of Washington, Seattle, WA 98195, USA}
\newcommand{\wuppertal}{Department of Physics, Faculty of Mathematics and Natural Sciences, University of Wuppertal, Gau{\ss}str.~20, 42119 Wuppertal, Germany}
\newcommand{\saclay}{IRFU (DPhP \& APC), CEA, Universit\'{e} Paris-Saclay, 91191 Gif-sur-Yvette, France }
\newcommand{\tum}{Technische Universit\"{a}t M\"{u}nchen, James-Franck-Str.~1, 85748 Garching, Germany}
\newcommand{\uhd}{Institute for Theoretical Astrophysics, University of Heidelberg, Albert-Ueberle-Str.~2, 69120 Heidelberg, Germany}
\newcommand{\tunl}{Triangle Universities Nuclear Laboratory, Durham, NC 27708, USA}
%
%
\newcommand{\ornl}{Also affiliated with Oak Ridge National Laboratory, Oak Ridge, TN 37831, USA}
%
%
%

\affiliation{\iap}
\affiliation{\ipe}
\affiliation{\inr}
\affiliation{\muenster}
\affiliation{\etp}
\affiliation{\itep}
\affiliation{\tum}
\affiliation{\mpp}
\affiliation{\unc}
\affiliation{\tunl}
\affiliation{\lbnl}
\affiliation{\wuppertal}
\affiliation{\madrid}
\affiliation{\washington}
\affiliation{\npi}
\affiliation{\massit}
\affiliation{\cmu}
\affiliation{\saclay}
\affiliation{\mpik}
\affiliation{\berlin}
\affiliation{\uhd}
\affiliation{\mainz}


\author{M.~Aker}\affiliation{\iap}
\author{D.~Batzler}\affiliation{\iap}
\author{A.~Beglarian}\affiliation{\ipe}
\author{J.~Behrens}\affiliation{\iap}
\author{A.~Berlev}\affiliation{\inr}
\author{U.~Besserer}\affiliation{\iap}
\author{B.~Bieringer}\affiliation{\muenster}
\author{F.~Block}\affiliation{\etp}
\author{S.~Bobien}\affiliation{\itep}
\author{B.~Bornschein}\affiliation{\iap}
% ...

\collaboration{KATRIN Collaboration}\noaffiliation

\title{Search for keV-scale Sterile Neutrinos with first KATRIN Data} 

\begin{abstract}
    In this work we present a keV-scale sterile-neutrino search with the first tritium data of the KATRIN experiment, acquired in the commissioning run in 2018. KATRIN performs a spectroscopic measurement of the tritium $\upbeta$-decay spectrum with the main goal of directly determining the effective electron anti-neutrino mass. During this commissioning phase a lower tritium activity facilitated the search for sterile neutrinos with a mass of up to $1.6\, \mathrm{keV}$. We do not find a signal and set an exclusion limit on the sterile-to-active mixing amplitude of down to  $\sin^2\theta < 5\cdot10^{-4}$ ($95\,\%$ C.L.), improving current laboratory-based bounds in the sterile-neutrino mass range between 0.1 and $1.0\, \mathrm{keV}$.
\end{abstract}

\maketitle


\newpage
\section{Introduction}
Right-handed neutrinos are a minimal and well-motivated extension of the Standard Model of Particle Physics (SM)~\cite{Adhikari:2016bei}. Right-handed neutrinos, as opposed to the known left-handed neutrinos, would not interact in any SM interaction and are therefore called sterile neutrinos. The introduction of right-handed partners to the left-handed neutrinos provides a natural way to create neutrino masses~\cite{eVWhitePaper}. No gauge symmetry of the SM forbids the introduction of a Majorana mass term of arbitrary scale for the right-handed neutrino. As a consequence, new neutrino-mass eigenstates arise, which are mostly sterile, but can have an admixture of the active SM neutrinos~\cite{Volkas:2001zb}. The size of the admixture is typically given by $\sin^2\theta$, where $\theta$ refers to the active-to-sterile mixing angle. In the following the new mass eigenstates are referred to as \textit{sterile neutrinos}.

Very light sterile neutrinos in the eV-mass range are motivated by long-standing anomalies in short-baseline-oscillation experiments~\cite{eVWhitePaper, Giunti:2019aiy, Boser:2019rta}. Sterile neutrinos in the keV scale are viable candidates for dark matter~\mbox{\cite{PhysRevLett.72.17, PhysRevLett.82.2832, Boyarsky:2018tvu, Adhikari:2016bei}}. For very large masses ($>$GeV), sterile neutrinos could solve the puzzle of the lightness of active neutrinos via the see-saw mechanism and may shed light on the matter/anti-matter asymmetry of the universe~\cite{Petcov:2013, Drewes:2016jae, PhysRevLett.110.061801}. 

A notable feature of sterile-neutrino dark matter is that it can act as effectively cold or warm dark matter depending on its production mechanism in the early universe~\cite{1475-7516-2017-11-046}. This property can help mitigate tensions between predictions of purely cold dark-matter scenarios and observations of small-scale structures in the universe. The existence of sterile-neutrino dark matter is strongly bound by indirect searches and cosmological observations, which limit their mixing amplitude with active neutrinos to $\sin^2\theta < 10^{-6} - 10^{-10}$ in a mass range of $(1 - 50) \, \mathrm{keV}$~\cite{limit1, limit2, limit3, limit4, limit5}. These limits can be model-dependent and could potentially be circumvented~\cite{Benso:2019jog}. Current laboratory-based limits are orders of magnitude weaker~\cite{Abdurashitov2017, Radcliffe:1992, PhysRevC.36.1504, PhysRevC.32.2215, OHI1985322, Ohsh93, Muel94, Holz99}. 

\begin{figure}[]
\centering
		\includegraphics[width=\linewidth]{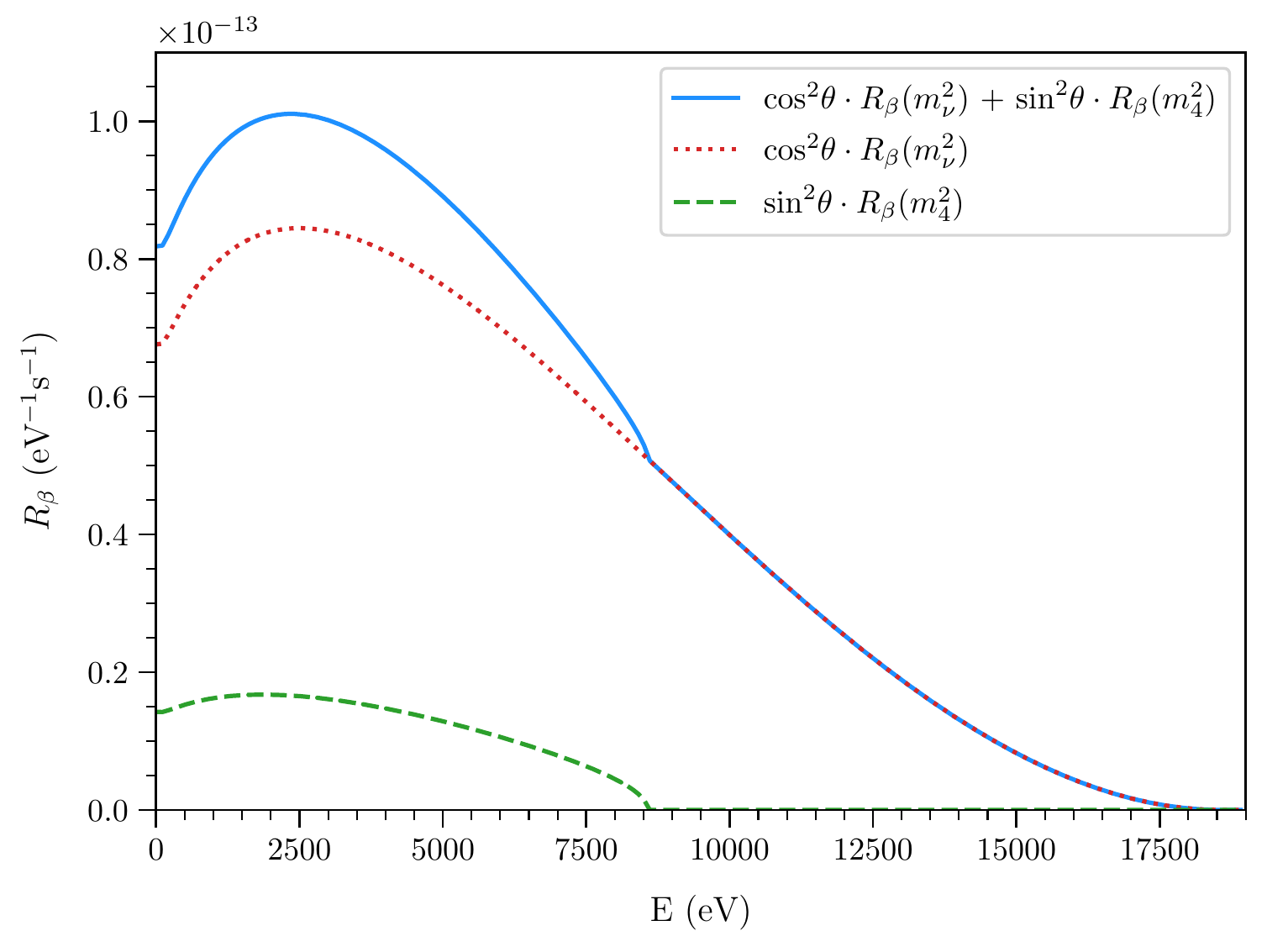}
		\label{Fig:Signal}
        \caption{Illustration of a keV-scale sterile-neutrino signature in the tritium $\upbeta$-decay spectrum. The position of the kink-like signal is determined by the mass of the sterile neutrino $m_4$ and the amplitude is governed by the mixing amplitude $\sin^2\theta$. The value for the mixing amplitude in the figure is unrealistically large, and was chosen for illustrative purpose. }
\end{figure}

Sterile neutrinos with masses $m_\mathrm{4} \lesssim E_0 = 18.6\, \mathrm{keV}$ are accessible in tritium $\upbeta$-decay~\cite{SHROCK1980159, Mer:2015a, Mer:2015b}, with $E_0$ being the kinematic endpoint, i.e. the maximum energy the electron can obtain for the case of zero neutrino mass. For electron energies of $E \leq E_0-m_\mathrm{4}$ the emission of a neutrino with the mass $m_\mathrm{4}$ along with the $\upbeta$-electron is kinematically allowed.  As a consequence, a sterile neutrino would manifest itself as a kink-like feature and spectral distortion at $E_\mathrm{kink}<E_0-m_\mathrm{4}$, as illustrated in figure~\ref{Fig:Signal}.

The Karlsruhe Tritium Neutrino experiment (\mbox{KATRIN})~\cite{KATRIN:2021dfa} has one of the strongest tritium sources used for scientific research. The goal of the experiment is to probe the effective electron anti-neutrino mass with a sensitivity of $0.2\,\mathrm{eV}$ at \SI{90}{\percent} confidence level after approximately 5 years of measurement time. This is achieved by analyzing the shape of the tritium $\upbeta$-decay spectrum near the endpoint at $E_0 = \SI{18.6}{\kilo\electronvolt}$, where the impact of the neutrino mass is maximal. KATRIN combines a high-luminosity tritium source with a high-resolution electrostatic filter of the MAC-E-filter (magnetic adiabatic collimation and electrostatic filter) type~\cite{Angrik:2005ep}. Recently, KATRIN published its first sub-eV limit on the effective electron anti-neutrino mass of $0.8\,\mathrm{eV}$ ($90\,\%$ CL)~\cite{KATRIN:KNM1PRL, KATRIN:KNM1Analysis, KATRIN:2021uub}, based on the first two high-tritium-activity data-taking campaigns. 


\begin{figure*}[]
\centering
		\includegraphics[width=\textwidth]{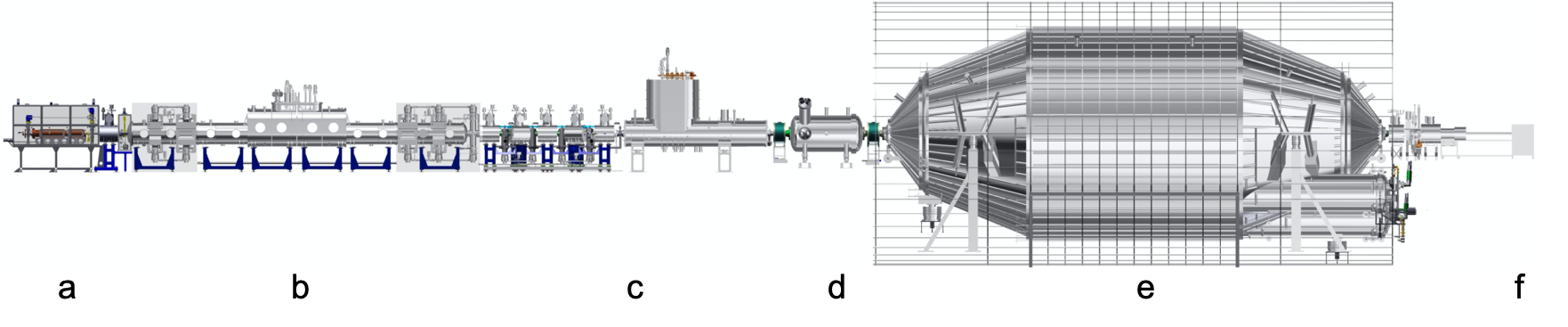}
		\label{Fig:KATRINsetup}
        \caption{The experimental setup of the 70-m-long KATRIN beamline. Gaseous molecular tritium is inserted through capillaries at the center of the Windowless Gaseous Tritium Source (WGTS) (b). $\upbeta$-electrons created in the 10-m long WGTS are guided with a system of superconducting solenoids through the transport section (c) towards the spectrometer section. The pre-spectrometer (d) can pre-filter electrons and the main spectrometer (e) transmits only electrons above a sharp adjustable transmission edge. The 148-pixel focal plane detector (f) counts the transmitted electrons as a function of the main spectrometer's transmission edge. Non-transmitted electrons are eventually absorbed in the rear wall of the rear-section (a) of the beam line.}
\end{figure*} 

Several studies~\cite{Mer:2015a, Mer:2015b, Huber2021_1000128344} have shown that a KATRIN-like measurement also provides a promising sensitivity to eV- and keV-scale sterile neutrinos. Based on the first two KATRIN measurement campaigns, improved limits could be set on eV-scale sterile neutrinos~\cite{KATRIN:2020dpx,KATRIN:2022ith}. While an eV-scale sterile neutrino leaves a signature within the standard measurement interval of KATRIN, which extends to about 40 - 100~eV below $E_0$, the signature of a keV-scale sterile neutrino lies further away from the endpoint, outside of this interval. Consequently, a search for keV-scale sterile neutrinos requires an extension of the measurement interval which bears several challenges. One of them is the fact that the count rates deep in the spectrum exceed the level that can be resolved by the KATRIN focal-plane detector system~\mbox{\cite{Wall201473, Amsbaugh201540}}. A novel detector system, the so-called \textsc{TRISTAN} detector, is under development and designed to overcome this limitation~\cite{Mertens:2018vuu}. On the other hand, at the cost of reduced statistics, it is also possible to extend the measurement interval by reducing the source activity~\cite{Huber2021_1000128344}.

In 2018, the KATRIN beamline was operated for the first time with a small amount of tritium gas~\cite{KATRIN:2019gru}. For safety reasons, the isotopic abundance of tritium in the deuterium carrier gas was set to only $0.5\,\%$ in this commissioning campaign. The reduced tritium activity provided a unique opportunity to explore the spectrum in a wide energy range down to $1.6\, \mathrm{keV}$ below the endpoint. The data set obtained in the 12-day-long series of measurements allowed us to perform a search for sterile neutrinos in the $0.01 - \SI{1.6}{keV}$ mass range, with competitive sensitivity compared to previous laboratory-based searches.


\section{The KATRIN experiment}
The KATRIN experiment consists of a 70-m-long beamline~(figure~\ref{Fig:KATRINsetup}), combining a gaseous molecular tritium source with a MAC-E filter spectrometer to obtain a high-precision, high-statistics integral $\upbeta$-decay spectrum. A detailed description of the apparatus can be found in~\cite{KATRIN:2021dfa}. 
 
The windowless gaseous tritium source (WGTS) consists of a $10\,\mathrm{m}$ long stainless-steel tube with a diameter of $90\,\mathrm{mm}$. Highly purified tritium gas is injected continuously at the center of the WGTS and diffuses to the up- and downstream end of the source tube where it is pumped out and fed back to the tritium loop system that is integrated in the infrastructure of the Tritium Laboratory Karlsruhe (TLK). 

The source and spectrometer sections of the KATRIN beamline are connected by the so-called transport section. Here, differential and cryogenic pumping sections reduce the tritium flow by more than 14 orders of magnitude, while the electrons are guided adiabatically to the spectrometers by a system of superconducting magnets. 

The high-resolution main spectrometer selects the electrons according to their energy, by applying the MAC-E filter technique. The MAC-E filter only transmits electrons with a longitudinal kinetic energy (kinetic energy component associated with the motion parallel to the magnetic field lines) larger than the retarding energy $qU$, where $U$ is the precisely adjustable voltage of the spectrometer~\cite{Rodenbeck:2022iys} and $q$ refers to the electron charge. A magnetic field, which decreases by approximately 4 orders of magnitude from the ends to the center of the spectrometer, transforms the total kinetic energy of the electrons into longitudinal energy. The MAC-E filter technology combines a large angle acceptance of $51^{\circ}$ with an energy resolution at the eV-scale.

Electrons that overcome the retarding potential in the main spectrometer are counted at the focal-plane detector (FPD). The FPD is a monolithic silicon array, radially and azimuthally segmented in 148 pixels~\cite{Wall201473, Amsbaugh201540}. By measuring the count rate at different retarding energies, the integral $\upbeta$-decay spectrum is obtained. In order to increase the signal-to-background ratio, the transmitted electrons are accelerated by a post-acceleration electrode (PAE) with an electrostatic potential of $U_\mathrm{PAE}=10 \, \mathrm{kV}$ before impinging on the detector surface. 

\section{The First Tritium Campaign}
The First Tritium (FT) campaign, which inaugurated the KATRIN experiment, was a commissioning campaign to demonstrate the stable operation of the integral system and test different analysis strategies. A technical description of the measurement campaign and the results with respect to stability and analysis techniques can be found in~\cite{KATRIN:2019gru}. 

\subsection{Tritium source operation}
During the FT campaign, the WGTS was operated at a column density (gas density integrated over the length of the source) of \mbox{$\rho d = 4.46\cdot10^{17} \, \mathrm{molecules}/\mathrm{cm}^2$}, with a reduced tritium activity of $500\,\mathrm{MBq}$, which corresponds to \SI{0.5}{\percent} of the activity used for neutrino-mass measurements. This activity limitation was achieved by mixing traces of tritium (in the form of DT) with pure deuterium ($\text{D}_2$) ~\cite{doi:10.1080/15361055.2019.1705681, doi:10.1080/15361055.2019.1668253}. This gas mixture was circulated through the WGTS via the main tritium loop~\cite{PRIESTER201542}. At all times, the gas composition was monitored by a Laser Raman spectroscopy (LARA) system~\cite{Sturm2010_1000019355, Schloesser2013_1000034967} and by the Forward Beam Monitor (FBM)~\cite{Beglarian:2021ubj}. In the FT experimental configuration the downstream end of the KATRIN beam line was terminated by a stainless-steel gate valve rather than the rear wall. 
\begin{figure}[t]
\centering
		\includegraphics[width=\linewidth]{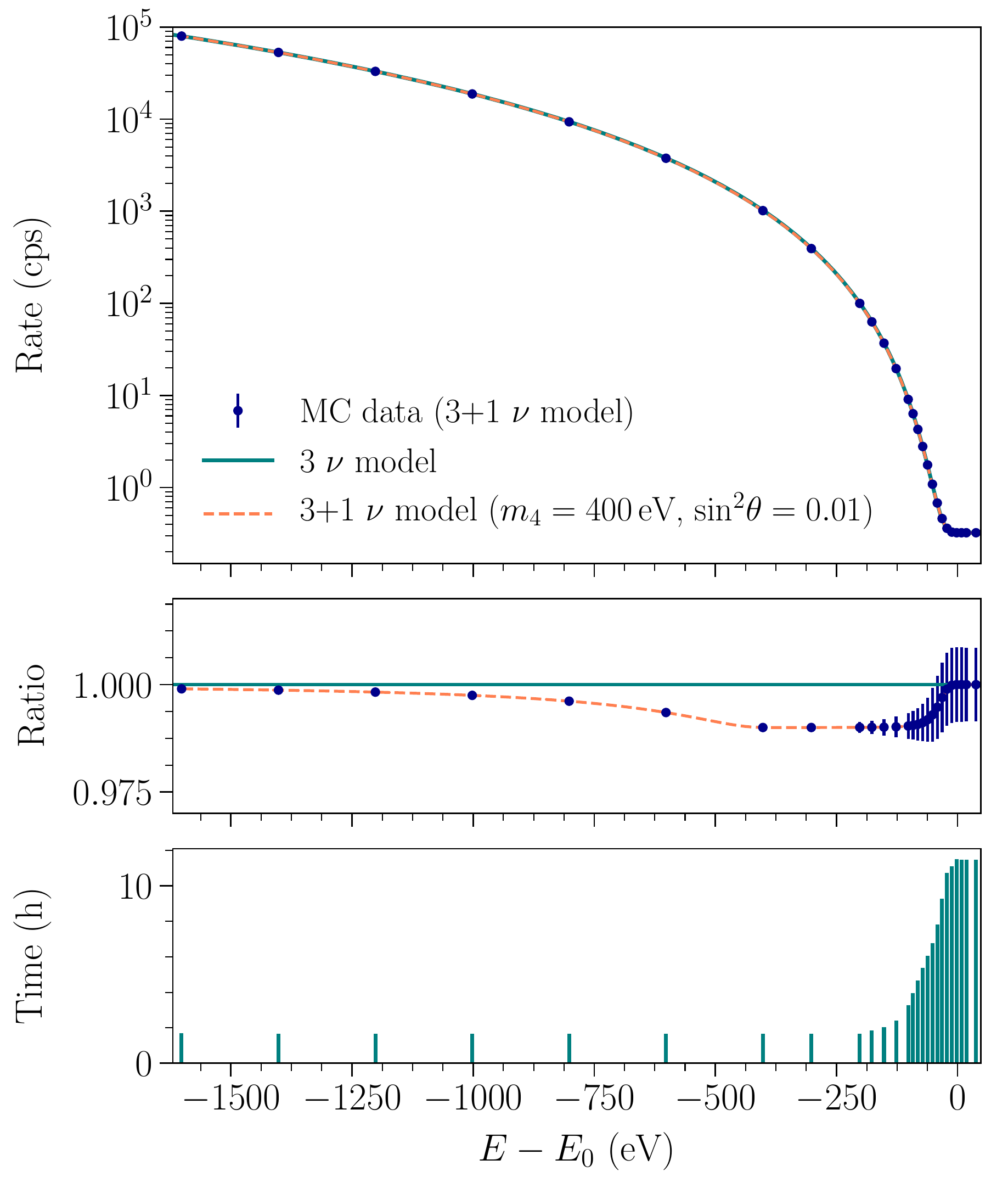}
		\label{Fig:signal}
        \caption{The top panel shows a simulated $\upbeta$-decay spectrum including a fourth, sterile mass eigenstate with $m_4=400\,\mathrm{eV}$ and a mixing amplitude of $\sin^2 \theta = 0.01$. The active neutrino mass is set to zero. The ratio to the standard $3\upnu$-model is displayed in the middle panel. At an energy of 400 eV below $E_{0}$ the additional sterile-neutrino branch kicks in and distorts the overall spectrum for energies $E < E_{0}-400\,\mathrm{eV}$. The bottom part of the figure shows the accumulated measurement time distribution of all analyzed tritium scans.}
\end{figure} 
\subsection{Spectrometer operation}
KATRIN obtains the integral $\upbeta$-decay spectrum in so-called scans, i.e. by sequentially applying different retarding energies $qU_i$ to the main spectrometer and counting the number of transmitted $\upbeta$-electrons $N(qU_i)$ with the focal plane detector. During the FT campaign, the spectrum was measured at 26 different retarding potentials in the range of $E_0 - \SI{1600}{\electronvolt} \leq qU_i \leq E_0 + \SI{30}{\electronvolt}$.  Figure~\ref{Fig:signal} shows the measurement-time distribution during FT data taking. The sequence of applied retarding potentials is either increasing (up scans) or decreasing (down scans). Applying up scans and down scans in an alternating fashion optimizes the averaging of possible drifts of slow-control parameters and also minimizes the time for setting the retarding potentials.

The FT measurement entails 122 scans 1 - 3 hours each, leading to a total measurement time of 168 hours. The $\upbeta$-decay spectrum obtained in each individual scan, was analyzed separately to test the stability of the system. The obtained effective endpoint of each spectrum shows an excellent stability, consistent with purely statistical fluctuations~\cite{KATRIN:2019gru}.

\section{Spectrum Calculation}
The expected integral $\upbeta$-decay spectrum is composed of two main parts: 1) the theoretical differential $\upbeta$-electron spectrum $R_\upbeta(E)$ and 2) the experimental response function $f_\mathrm{calc}(E, qU_i)$. The total calculated rate $R_\mathrm{calc}(qU_i)$ at a given retarding energy $qU_i$ is given by  
\begin{equation}
\label{Ntheo}
R_\mathrm{calc}(qU_i) = A_{\mathrm{s}} N_{\mathrm{T}} \int_{qU_i}^{E_0}R_\upbeta(E) f_\mathrm{calc}(E, qU_i) \ \dd E + R_{\mathrm{bg}},
\end{equation}
where $N_{\mathrm{T}}$ is the signal normalization, which includes the number of tritium atoms in the source, the maximum acceptance angle of the MAC-E filter and the detection efficiency. $A_{\mathrm{s}}$ is a free parameter in the fit and $R_{\mathrm{bg}}$ denotes the retarding-potential-independent background rate~\cite{Trost2019_1000090450}.

\subsection{Differential $\upbeta$-decay spectrum} 
Generally, the $\upbeta$-decay spectrum is a superposition of spectra corresponding to the different neutrino-mass eigenstates with masses $m_i$ that contribute to the electron flavor. Due to the tiny mass differences of the three light neutrino-mass eigenstates this superposition cannot be resolved with current experiments. However, a heavy, mostly sterile mass state, would lead to a distinct distortion of the total $\upbeta$-decay spectrum
\begin{equation}
R_\upbeta(E) = \cos^2\theta \, R_\upbeta(E, m_\upnu^2) + \sin^2\theta \, R_\upbeta(E,  m_4^2)
\end{equation}
where $R_\upbeta(E, m_\upnu^2)$ and $R_\upbeta(E,  m_4^2)$ correspond to the active and sterile decay branches, respectively. $\theta$ denotes the active-sterile neutrino mixing angle. $m_\upnu^2 = \sum_{i=1}^3 |U_{\upnu i}|^2 m_i^2$ is the squared effective electron anti-neutrino mass, where $U_{\upnu i}$ denote elements of the Pontecorvo–Maki–Nakagawa–Sakata matrix. $E$ denotes the kinetic energy of the $\upbeta$-electron.

The $\upbeta$ spectrum $R_\upbeta(E)$ of molecular tritium is described by Fermi's theory
\begin{widetext}
\begin{equation}
    R_\upbeta(E) = C \cdot F(Z',E) \cdot (E+m_\mathrm{e}) \cdot p \cdot \sum_i P_i\cdot (E_0-E-E_i)^2\cdot \sqrt{1-\biggl(\frac{m_\upnu}{E_0-E-E_i}\biggr)^2},
    \label{equ:fermi_theory}
\end{equation}
\end{widetext}
where $C= \frac{G_F^2}{2\pi^3}\cos^2\Theta_C |M_{\text{nucl}}|^2$ with $G_{\mathrm{F}}$ denoting the Fermi constant, $\Theta_{\mathrm{C}}$ the Cabibbo angle, and $M_{\text{nucl}}$ the energy-independent nuclear matrix element. The $F(E, Z')$ represents the Fermi function with $Z'=2$ for the atomic number of helium, the daughter nucleus in this decay. $E$, $p$, and $m_{\text{e}}$ denote the kinetic energy, momentum, and mass of the $\upbeta$-electron, respectively. 

After the $\upbeta$-decay of tritium in a DT molecule, the daughter molecule $\mathrm{^3HeD^+}$ can end up in an electronic ground state or excited state, each of which is broadened by rotational and vibrational excitations of the molecule~\cite{PhysRevC.91.035505}. As a consequence, this excitation energy $E_i$ reduces the available kinetic energy for the electron. Thus the differential $\upbeta$-electron spectrum is a superposition of spectra, corresponding to all possible final states, weighted by the probability $P_i$ for decaying into a certain final state $i$. For this analysis, we use the latest calculation of Saenz et al. for the isotopologue DT~\cite{PhysRevLett.84.242}.

The molecular final-state distribution depends slightly on the $\upbeta$-decay energy. Mainly, the mean and width of the ground-state distribution depends on the recoil energy of the daughter molecule, which in turn depends on the $\upbeta$-decay energy~\cite{PhysRevC.91.035505}. By taking into account this energy dependence in the theoretical calculation of the integral $\upbeta$-decay spectrum, we found that $R_\mathrm{calc}(qU_i)$ is altered by less than \SI{0.007}{\percent} for all retarding energies. Hence, we neglect the energy dependence of the final-state distribution in this analysis. 
Doppler broadening due to the thermal motion of tritium molecules in the source, which is operated at 30 K, is emulated as a broadening of the molecular final-state distribution~\cite{KleesiekBehrensDrexlin2019_1000092672}. 

\subsection{Response function}
The experimental response function 
\begin{widetext}
\begin{equation}
 f_\mathrm{calc}(E, qU_i) = \int_0^E T(E-\epsilon, qU_i) \left(P_0\,\delta(\epsilon) + P_1\,f(\epsilon) + \right.\\ 
 \left.P_2\,(f\otimes f)(\epsilon) + ...\right) \, \dd \epsilon,\label{eq:response}
\end{equation}
\end{widetext}
is the probability of an electron with a starting energy $E$ to reach the detector. It combines the transmission function $T$ of the MAC-E filter and the electron's energy losses $\epsilon$ in the source. The transmission function $T$ reflects the resolution of the main spectrometer and is governed by the magnetic fields at the starting position of the electron, the maximum field in the beamline, and the magnetic field in the spectrometer's analyzing plane. Energy losses due to inelastic scattering with the deuterium molecules in the source are described by the product of the $s$-fold scattering probabilities $P_s$ and the energy-loss function $f(\epsilon)$ convolved $(s-1)$ times with itself (denoted by $...$). We consider an energy-dependent cross-section, but treat the energy-loss function $f(\epsilon)$ as energy independent. Here we use an energy-loss function measured in situ for deuterium~\cite{KATRIN:2021rqj}. Synchrotron energy losses of $\upbeta$-electrons in the high magnetic field in the source and transport section are included as a correction to the transmission function. Furthermore, the response function is slightly modified due to the dependence of the path length (and therefore effective column density) on the pitch angle of the $\upbeta$-electrons~\cite{KleesiekBehrensDrexlin2019_1000092672}. 

\subsection{Wide-interval corrections}
\label{ssec:widerange}
Beyond the tritium spectrum calculation described above, we investigate specific effects relevant at energies further away from the endpoint, outside the nominal KATRIN analysis window. 

\subsubsection{Detection Efficiency} The total detection efficiency is of minor relevance as it only affects the normalization of the measured spectrum and not its shape. In contrast, a retarding-potential-dependent detection efficiency alters the shape of the integral spectrum. Figure~\ref{fig:corrections}$\,\mathrm{a}$ displays the retarding-potential dependence of the detection efficiency. It includes the following effects:

\paragraph{Region-of-interest coverage} In order to count the events at a given retarding potential, the measured rate at the focal-plane detector is integrated in a wide and asymmetric region of interest (ROI) of \mbox{$\SI{14}{\kilo\electronvolt} \leq E + qU_{\text{PAE}} \leq \SI{32}{\kilo\electronvolt}$}, where $E$ is the $\upbeta$-electron energy and \mbox{$U_{\text{PAE}}=\SI{10}{\kilo\electronvolt}$} is the post-acceleration voltage. This ROI is chosen to account for the moderate energy resolution of about \SI{3}{\kilo\electronvolt} (full-width-half-maximum) and the low-energy tail of the spectrum due to the energy loss of electrons in the dead layer and backscattering from the detector surface~\cite{Amsbaugh201540}. The same ROI is used for each retarding-potential setting. As the mean of the electron peak shifts with the retarding potential, some electrons move out of the fixed ROI, which effectively changes the detection efficiency. This change of detection efficiency is experimentally determined based on reference measurements, and is corrected accordingly~\cite{Leo:2019}. For this effect, we interpret the variation of the correction for all detector pixels used in the analysis as the uncertainty. Assuming a detection efficiency of $\epsilon_{\text{roi}}=1$ at $E_0$, we find a relative detection efficiency at $1\,\mathrm{keV}$ below $E_0$ of $\epsilon_{\text{roi}}=0.99911 \pm 0.00036$. 

\paragraph{Pile-up} As the counting rate at the focal-plane detector depends on the retarding potential, so does the probability of pile-up. Most pile-up events occur outside the ROI, thereby effectively changing the detector efficiency~\cite{Korzeczek2020_1000120634}. We estimate the detection efficiency $\epsilon_{\text{pu}}$ with a two-fold random coincident model, according to
\begin{equation}
    \epsilon_{\text{pu}}(R)= (1-\frac{\alpha}{2})\exp{-2WR}+\frac{\alpha}{2},
\end{equation}
where $R$ is the Poissonian-distributed signal rate, $1-\frac{\alpha}{2} = 0.79\pm 0.02$ denotes the pile-up event rejection ratio, and $W = 1.826 \pm 0.026 $~ns denotes the effective window length of the trapezoidal energy filter used to determine the energy of each event~\cite{Amsbaugh201540}. The uncertainty of this correction is determined by the uncertainty of these model parameters. At $1\,\mathrm{keV}$ below $E_0$, pile-up reduces the detector efficiency to $\epsilon_{\text{pu}} = 0.99952 \pm 0.00001$. 
    
\paragraph{Backscattering} A significant fraction of about \SI{20}{\percent} of all electrons impinging on the detector surface are backscattered. For low retarding potentials and small energy depositions in the detector, these backscattered electrons have a chance of getting lost by overcoming the retarding potential a second time. The lower the retarding potential, the higher is the probability to lose electrons this way, effectively changing the detection efficiency~\cite{Korzeczek2020_1000120634}. We estimate this effect by Monte Carlo simulations with the KATRIN-specific simulation packages \textsc{Kess}~\cite{Renschler2011_1000024959} and \textsc{Kassiopeia}~\cite{kassiopeia}.

We estimate the uncertainty of this correction by changing the input parameters according to their uncertainties. The relevant parameters here are the magnetic field at the position of the detector $B_\mathrm{det}$ and the maximal magnetic field upstream of the detector magnet in the pinch magnet $B_\mathrm{pch}$. The uncertainties on the magnetic fields are estimated via by comparisons of measurements and simulations~\cite{KATRIN:2019gru} and are quoted in table~\ref{tab:systematics}. Moreover, as the Si-crystal lattice orientation relative to the electron's incident angle is not precisely known, we allow for an uncertainty of the amplitude of the elastic backscattering peak. We conservatively vary the amplitude obtained by Monte-Carlo simulations by \SI{+50}{\percent}, emulating the two extreme cases of anomalous transmission and absorption~\cite{elasticbs}. At $1\,\mathrm{keV}$ below $E_0$, backscattering reduces the detector efficiency to $\epsilon_{\text{bs}} = 0.99893 \pm 0.00027$. 

\subsubsection{Rear-wall backscattering} Another effect which is negligible in the case of an endpoint analysis is the detection of $\upbeta$-electrons which are backscattered at the rear wall of the beamline and still reach the focal-plane detector. During the FT measurement campaign a stainless-steel gate valve terminated the beamline. In the backscattering process, the electrons lose some amount of energy, which typically forbids them to be transmitted through the main spectrometer. However, for low retarding potentials, there is a non-negligible probability for this transmission to occur~\cite{Huber2021_1000128344}. The backscattering of tritium $\upbeta$-decay electrons from the stainless-steal plate was simulated with \textsc{GEANT4}, providing the backscattering probability as well as the energy and angle distribution of backscattered electrons. The corresponding correction to the integral $\upbeta$-decay spectrum is depicted in figure~\ref{fig:corrections}$\,\mathrm{a}$. 

We estimate the uncertainty of this correction by varying in the simulation the magnetic fields at the rear wall $B_\mathrm{rw}$ and in the source section $B_\mathrm{s}$ by their respective uncertainties given in table~\ref{tab:systematics}. In addition, we estimate a theoretical uncertainty arising from the \textsc{GEANT4} simulation, by computing the correction with different physics packages (i.e.\ the \textit{emlivermore} and \textit{emstandardSS} packages) and interpreting the difference as a measure of the uncertainty. At $1\,\mathrm{keV}$ below $E_0$, we find a multiplicative correction to the observed rate by $\epsilon_{\text{rw}} = 1.00097 \pm 0.00096$.

\subsubsection{Magnetic Trapping} 
\begin{figure*}[]
    \subfigure[]{\includegraphics[width=0.49\textwidth]{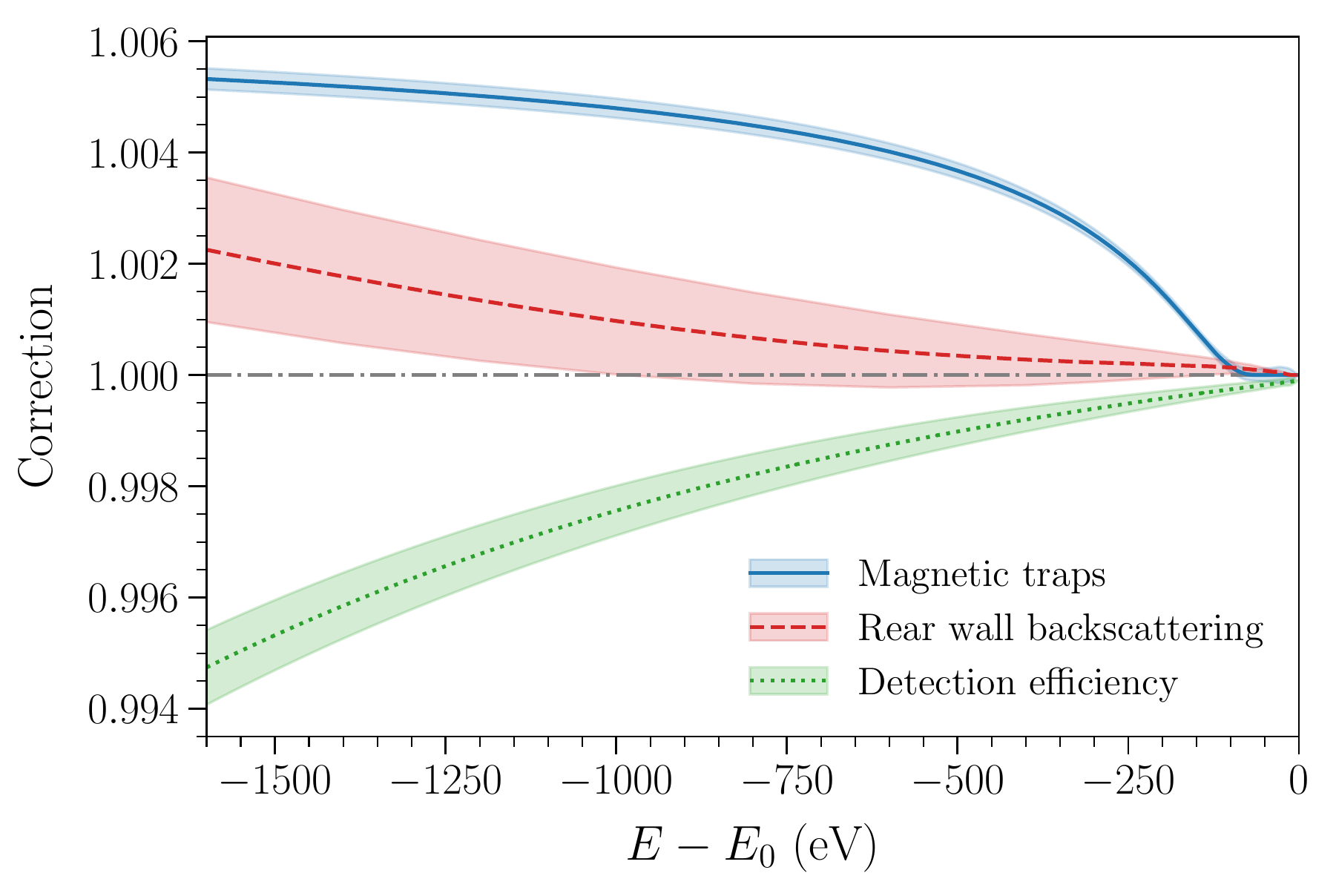}}
    \subfigure[]{\includegraphics[width=0.49\textwidth]{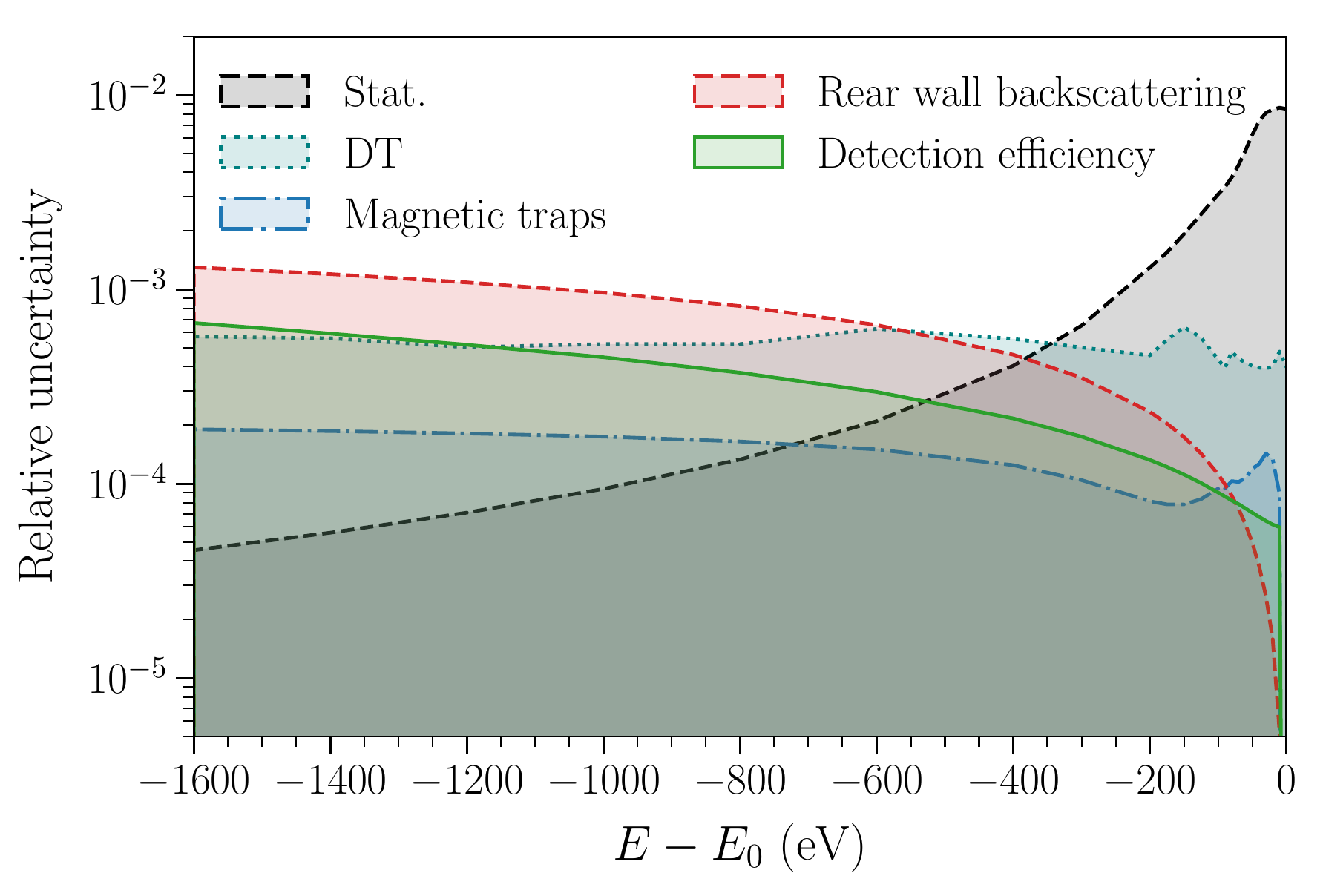}}
    \caption{a) Spectral corrections due to retarding-potential-dependent detection efficiency (green), magnetic trapping of $\upbeta$-electrons in the WGTS (blue), and backscattering of $\upbeta$-electrons on the rear wall of the WGTS (red). b) Relative Poisson statistical uncertainty of the spectral data points (grey), relative statistical uncertainty arising from the deuterium-tritium (DT) source activity fluctuations (light blue), relative spectral uncertainties arising from the three corrections displayed in a) (blue, red, green). Note that the latter three uncertainties are correlated between the spectral data points.}
    \label{fig:corrections}
\end{figure*}

The source beam line exhibits small local magnetic field minima, arising from the small gaps between adjacent superconducting coil units. Electrons starting with a pitch angle larger than a certain threshold in such local magnetic field minima can be magnetically trapped. Frequent elastic and inelastic scattering change their angle and they eventually escape from the trap with reduced energy. If the retarding potential of the spectrometer is low enough, these electrons have a chance to reach the detector~\cite{Huber2021_1000128344}. Based on a Monte Carlo simulation with \textsc{Kassiopeia}, we calculate the corresponding correction to the integral $\upbeta$-decay spectrum, as displayed in figure~\ref{fig:corrections}$\,\mathrm{a}$.  

We obtain the uncertainty on the correction by varying the relevant simulation input parameters, namely the source and pinch magnetic field $B_\mathrm{s}$ and $B_\mathrm{pch}$, the gas density in the source $\rho$, and the parameters of the energy loss function. At $1\,\mathrm{keV}$ below $E_0$, we find a multiplicative correction to the observed rate by $\epsilon_{\text{mt}} = 1.00510 \pm 0.00017$.


\subsubsection{Non-adiabaticity} At low retarding potentials of the MAC-E filter, some electrons have a comparatively high surplus energy. This is of concern, since the magnetic guiding field drops from about \SI{5}{\tesla}, at the entrance to about $6\cdot10^{-4}$~T in the center of the spectrometer. If an electron experiences an excessive change of the magnetic field within one cyclotron circle, it exhibits non-adiabatic motion. The non-adiabatic motion causes a chaotic change of the pitch angle and hence a possible magnetic reflection at the exit of the spectrometer. Eventually this can lead to a reduction of the number of transmitted electrons~\cite{Huber2021_1000128344}. A full Monte Carlo simulation with \textsc{Kassiopeia} shows that in the realistic magnetic field settings of the FT campaign, non-adiabatic effects can indeed occur at more than $1\,\mathrm{keV}$ below the endpoint. However, averaged over all radii in the spectrometer, this effect leads only to a small reduction of the rate of less than $0.01\,\%$ for all retarding potentials used in this measurement and can thus be neglected.

\section{Data selection and combination}
The full FT data set is sub-divided into several parts: 1) As mentioned above, the integral tritium spectrum is recorded in 122 scans to accommodate temporal changes of slow-control parameters, such as the source activity. 2) Each of the 148 pixels of the focal-plane detector measures a statistically independent tritium $\upbeta$-decay spectrum, to take into account radial and azimuthal variations of the electric and magnetic fields in the analyzing plane. 

For this analysis, we combine a selection of 82 ``golden'' scans by adding the counts recorded at each retarding potential set point, called \textit{scan step}, to construct a high-statistics single spectrum with $n_{\text{scan-step}}=26$ data points. Equivalently, we combine 119 ``golden'' pixels in a single effective pixel, by adding all counts and assuming an average response function for the entire detector. Simulations have shown that these assumptions lead to a negligible error on the fitted parameters~\cite{KATRIN:2019gru}. We exclude scans that were performed at at different experimental settings, such as at a different column density or with different HV set poionts. We exclude pixels which do not record the full flux of electrons due to misalignment. A full description of the data quality criteria can be found in~\cite{KATRIN:2019gru}.

\section{Method of exclusion limit construction}
\label{sec:exclLimConstr}
The calculated model spectrum $\vec{R}_\mathrm{calc}$ is fit to the data $\vec{R}_\mathrm{data}$ by minimizing
\begin{equation}
\label{eq:chi2}
    \chi^2(\theta) = (\vec{R}_\mathrm{calc}(\alpha_i)-\vec{R}_\mathrm{data})^{\mathsf{T}} C^{-1} (\vec{R}_\mathrm{calc}(\alpha_i)-\vec{R}_\mathrm{data}),
\end{equation}
with respect to the nuisance parameters $\alpha_i$, while keeping the sterile neutrino mass $m_4$ and mixing amplitude $\sin^2\theta$ fixed at a given value. The nuisance parameters in this analysis are the signal normalization, the effective endpoint of the spectrum, and an overall background rate. $C$ is the covariance matrix, which contains both statistical and systematic uncertainties. The fit is repeated on a fine grid of fixed tuples ($m_4$, $\sin^2\theta$). According to Wilks' theorem~\cite{Wilks}, the $95\,\%$ confidence level (C.L.) exclusion limit is constructed by determining the $\Delta \chi^2=\chi^{2}(m_4, \sin^2\theta)-\chi^2_{\mathrm{min}} < 5.99$ contour, where $\chi^2_{\mathrm{min}}$ corresponds to the global best fit. The applicability of Wilk's theorem was tested with Monte-Carlo simulations for the null hypothesis.

\begin{table*}
	\caption{Summary of systematic uncertainties. We lists the 1-$\sigma$ uncertainties of the input parameters used to construct the covariance matrices. The energy-loss function is described with an empirical model, consisting of three Gaussian functions, which parameters are correlated. The correlation is not shown in the table.  }
	\label{tab:systematics}
	\centering
	\begin{tabular}{p{5cm}p{7cm}R{5cm}} \\
	\hline 
	                         Effect &  Description & Uncertainty (1$\sigma$)\\
	                   
    \hline
    \hline 
	\multirow[t]{2}{*}{Source scattering} & column density    &   \SI{3}{\percent} \\
	                                      & inel. scat. cross-section      &   \SI{2}{\percent} \\
    \hline
    \multirow[t]{5}{*}{Energy-loss function}       & normalizations $A_1$, $A_2$, $A_3$ &  \SI{6.14}{\percent}, \SI{0.47}{\percent}, \SI{0.65}{\percent}   \\
                                                   & means $\mu_1$, $\mu_2$, $\mu_3$& \SI{0.15}{\percent}, \SI{0.03}{\percent}, \SI{0.05}{\percent}   \\
                                                   & standard deviations $\sigma_1$, $\sigma_2$, $\sigma_3$ &  \SI{7.58}{\percent}, \SI{0.81}{\percent}, \SI{2.62}{\percent}  \\
	\hline
	\multirow[t]{3}{*}{Final-state distribution} &  normalization                   & \SI{1}{\percent}   \\
	                                             &  ground-state variance           & \SI{1}{\percent}   \\
	                                             &  excited-states variance         & \SI{3}{\percent}   \\
	\hline
	\multirow[t]{3}{*}{Magnetic fields}     &  source $B_\mathrm{s}$            & \SI{2.5}{\percent}  \\
                          	                &  analyzing plane $B_\mathrm{rw}$   & \SI{4.4}{\percent}\\
                          	                &  analyzing plane $B_\mathrm{ana}$   & \SI{1}{\percent}   \\
                                            &  maximum field at pinch $B_\mathrm{pch}$             & \SI{0.2}{\percent}  \\
	\hline
	Background          &  retarding-potential dependence              & \SI{5}{\milli cps\per\kilo\electronvolt}     \\
    \hline
    \multirow[t]{4}{*}{DT activity fluctuation}        &  uncorrelated &  \SI{0.05}{\percent} \\	
    \hline
    \multirow[t]{3}{*}{Detection efficiency}      & pixel variation   & \SI{0.2}{\percent}\\
                                                  & pile-up rejection fraction & \SI{2}{\percent}  \\
                                                  & energy-filter window length &  \SI{1.4}{\percent} \\
                                                  & elastic backscattering amplitude &  \SI{50}{\percent} \\
    \hline
    \multirow[t]{3}{*}{Rear-wall backscattering} &  difference of the mean backscattering probability between two GEANT-4 libraries & \SI{1.4}{\percent} \\
    \hline
    Non-adiabaticity & neglected & \\
    \hline
    Energy-dependence of FSD & neglected & \\
    \hline
	\end{tabular}
\end{table*}

\section{Systematic uncertainties}
\label{sec:systematic_uncertainties}
To include systematic uncertainties, the so-called covariance-matrix method is applied~\cite{Lisa:2019}. Here, the spectrum prediction is computed about $10^4$ times while varying the systematic parameters according to a Gaussian distribution, which width corresponds to the 1-$\sigma$ uncertainty of each parameter. In this way, the variance and also the covariance of the spectral data points $\vec{R}_\mathrm{calc}$, caused by the uncertainty of the systematic parameter, are extracted. The covariance matrix, $C$, is then included in the $\chi^2$-function, as can be seen in equation~(\ref{eq:chi2}).

\begin{figure}[]
\centering
		\includegraphics[width=\linewidth]{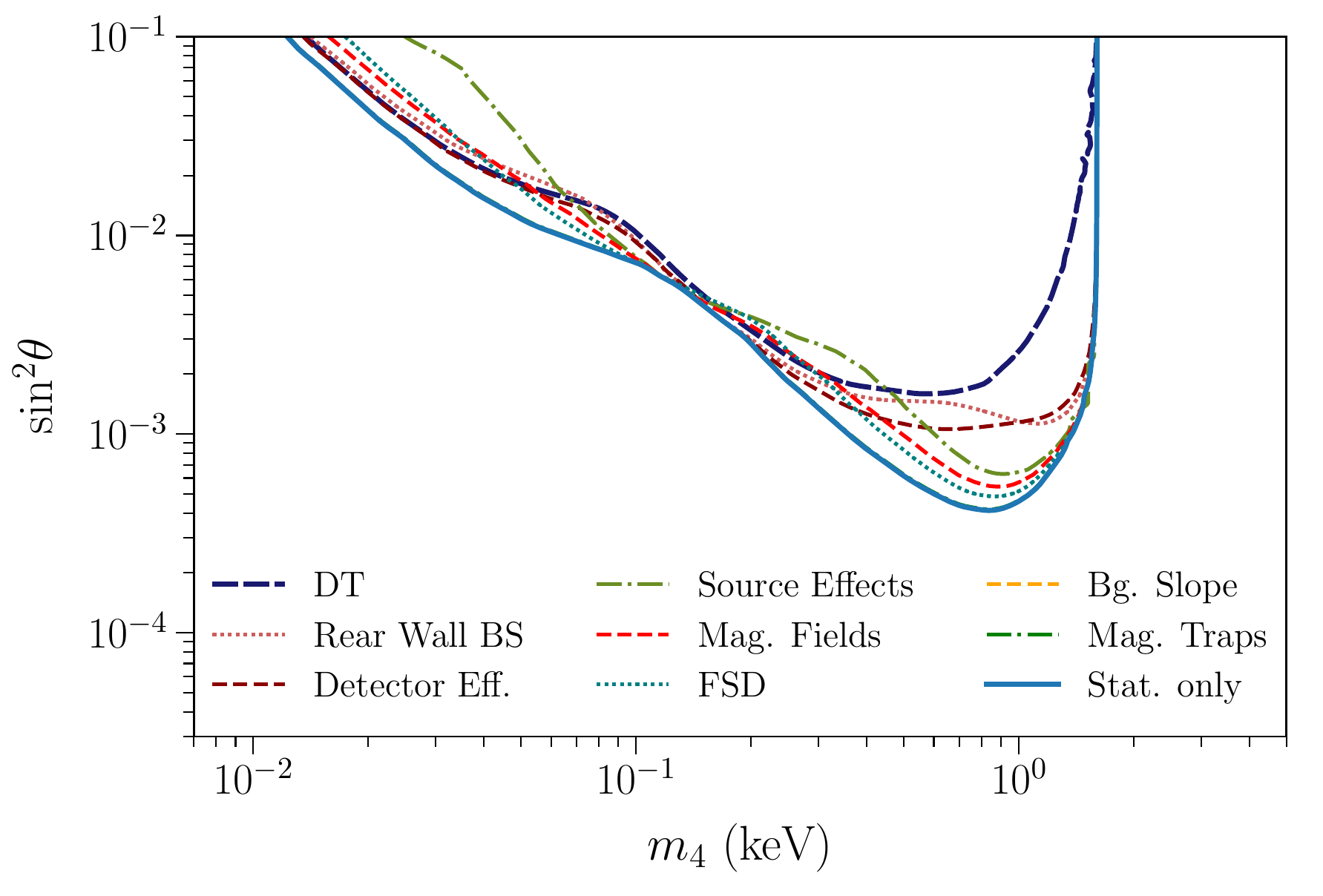}
        \caption{$95\,\%$ confidence level (C.L.) sensitivity to sterile neutrinos based on a Monte-Carlo copy of the first tritium data set. The statistical-only sensitivity is displayed by the solid blue line. The dashed lines show the impact of the statistical and individual systematic uncertainties.}
        \label{fig:sensitivity_systematics}
\end{figure}

We consider the standard KATRIN systematic uncertainties, described in detail in~\cite{KATRIN:2019gru}, and uncertainties arising from the wide-range corrections, described in section~\ref{ssec:widerange}. All systematic uncertainties are listed in table~\ref{tab:systematics}. Figure~\ref{fig:sensitivity_systematics} displays the impact of the individual systematic uncertainties on the $95\,\%$ C.L. sensitivity of a sterile-neutrino search, based on a Monte Carlo simulation of the FT data set. The following list briefly summarizes the effects, starting with those that have the largest impact on the sensitivity. 
\begin{itemize}
    \item Activity fluctuations: The uncertainty of the tritium activity in each scan step acts as an additional statistical error. With a relative magnitude of $5\cdot10^{-4}$ it dominates over the Poisson error (arising from the counting statistics) for retarding energies of $qU < E_0 - 400\, \mathrm{eV}$. This uncertainty is dominated by the statistical uncertainty of the LARA and FBM systems. Accordingly, this error will be reduced when operating at higher activity and with longer measurement time. Given its statistical nature, it is the most limiting uncertainty in this analysis, as can be seen in figure~\ref{fig:sensitivity_systematics}.
    
    \item Rear-wall backscattering: The spectral uncertainty arising from electrons that scatter off the rear wall and then reach the detector amounts to $1\cdot10^{-3}$ at $qU < E_0-1000\, \mathrm{eV}$. As described in section~\ref{ssec:widerange}, this uncertainty is estimated by sampling from two GEANT-4~\cite{GEANT4:2002zbu} simulations with different physics packages according to a binomial distribution and simultaneously varying the rear wall and source magnetic fields. Even though the magnitude of this uncertainty is larger than the one from activity fluctuations, its impact on the sterile-neutrino sensitivity is smaller, as can be seen in figure~\ref{fig:sensitivity_systematics}. This is due to the fact that these uncertainties are strongly correlated between the different spectral data points, thus preventing this correction from mimicking a kink-like sterile-neutrino signature~\cite{Mer:2015a}.
    
    \item Retarding-potential-dependent detector efficiency: The dominant contribution among the detector-related effects is the ROI coverage, as described in section~\ref{ssec:widerange}. In this case, the covariance matrix is constructed by sampling from the experimentally-determined efficiency corrections for different detector pixels, which vary by about $0.2\,\%$. The total uncertainty arising from the detector efficiency is of similar size to the activity fluctuations. However, this uncertainty is strongly correlated between the data points, and thus its impact on the sterile-neutrino sensitivity is mitigated, as can be seen in figure ~\ref{fig:sensitivity_systematics}.
    
    \item Source-scattering effects: Relevant parameters to describe energy losses due to scattering in the source section are the column density, the cross section, and the parameterized energy-loss function. The parameters and correlated uncertainties are determined via calibration measurements~\cite{KATRIN:2021rqj}. The covariance matrix is constructed by varying these parameters according to their uncertainties. The impact on the sterile-neutrino sensitivity is relatively large as the uncertainty on the column density during the FT campaign was rather high, since a calibration of the absolute column density was not available at that time.  
    
    \item Magnetic fields: The various magnetic fields of the KATRIN beamline are determined via a combination of magnetic-field measurements~\cite{KATRIN:Magnets,KATRIN:MOPS} and  simulations~\cite{KATRIN:FieldSimulation}. The source magnetic field shows the largest discrepancy between measurement and simulation of $2.5\,\%$. For the construction of the covariance matrix this difference between measurement and simulation is treated as a 1-$\sigma$ Gaussian uncertainty. Given this large uncertainty during the FT campaign, the sterile-neutrino sensitivity is visibly reduced, as can be seen in figure~\ref{fig:sensitivity_systematics}. With the help of new calibration methods, the magnetic-field uncertainties were reduced by up to one order of magnitude in later KATRIN measurement campaigns.
    
    \item Final-state distribution: Here we assume an uncertainty on the order of \SI{1}{\percent} on the probability to decay into the electronic ground-state and the broadening due to rotational and vibrational states. This uncertainty is larger than what is stated by experts on the theoretical calculations~\cite{Sae:2000} and it was chosen as a conservative estimation~\cite{KATRIN:KNM1Analysis}. In order to construct the covariance matrix, the individual final-state probabilities are varied according to these uncertainties. The conservatively large uncertainties on the final-state distribution lead to a small reduction of the sterile-neutrino sensitivity.
    
    \item The overall energy-independent background is treated as a free parameter in the fit. Additionally we allow for a retarding-potential dependence of the background. To this end, we introduce a so-called background slope, which is constrained by an external measurement to less than 5 mcps/keV. We find that this uncertainty leads to a negligible impact on the sterile neutrino search.
    
    \item The uncertainty arising from initially trapped electrons in local magnetic field minima the source section leads to a negligible effect on the sterile-neutrino sensitivity. 
    
\end{itemize}



\section{Results}
\begin{figure}[]
\centering
		\includegraphics[width=\linewidth]{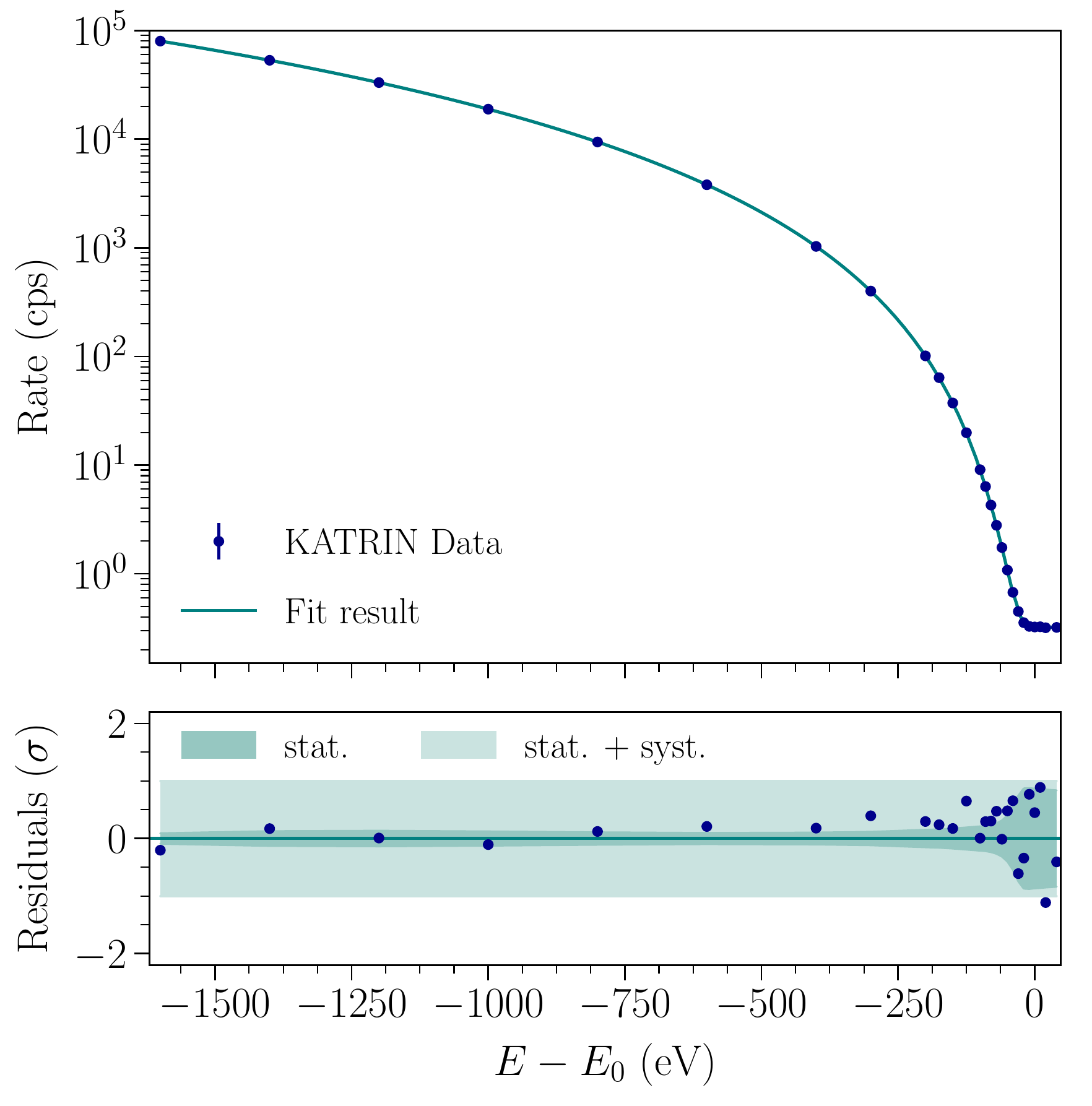}
        \caption{The best fit of all 82 spectra combined (by adding the counts at each retarding energy) with the normalized residuals expressed in standard deviation $\sigma$. The best fit is found for $m_4 = \SI{71.2}{\electronvolt}$ and $\sin^2\theta = 0.017$ with a goodness-of-fit of $\chi^2/\mathrm{ndof}=14.79/21$, and a corresponding p-value of 0.83.}
        \label{fig:fit}
\end{figure}

The statistics of the full data set amount to \SI{1.2E9}{} $\upbeta$-electrons. The corresponding spectrum with a fit including a sterile neutrino and all systematic uncertainties, shows an excellent agreement of the model with the data, with $\chi^2/\mathrm{ndof}=14.79/21$, and a corresponding p-value of 0.83, as shown in figure~\ref{fig:fit}.

As a first step we perform a sensitivity study based on Monte-Carlo generated data to assess the impact of the individual systematic effects, described in detail in section \ref{sec:systematic_uncertainties}. Figure~\ref{fig:sensitivity_systematics} displays the statistical sensitivity at the \SI{95}{\percent} C.L. and the sensitivities when including individual systematic uncertainties. The \SI{95}{\percent} C.L. statistical sensitivity reaches down to a value of $\sin^2\theta < 5\cdot10^{-4}$ at $m_4 = \SI{1000}{\electronvolt}$. Including all systematic uncertainties the best sensitivity is reduced to $\sin^2\theta < 2\cdot10^{-3}$. As discussed in detail in section~\ref{sec:systematic_uncertainties}, we find the DT activity fluctuation to be the dominant uncertainty. This uncertainty is reduced in future campaigns by increasing the total number of electrons collected at a given retarding potential, which is achieved with longer measurement times and higher source activities. Moreover, source-activity fluctuations would play a minor role in a differential measurement of the tritium $\upbeta$-decay spectrum, as planned with the TRISTAN detector. 

\begin{figure}[]
\centering
		\includegraphics[width=\linewidth]{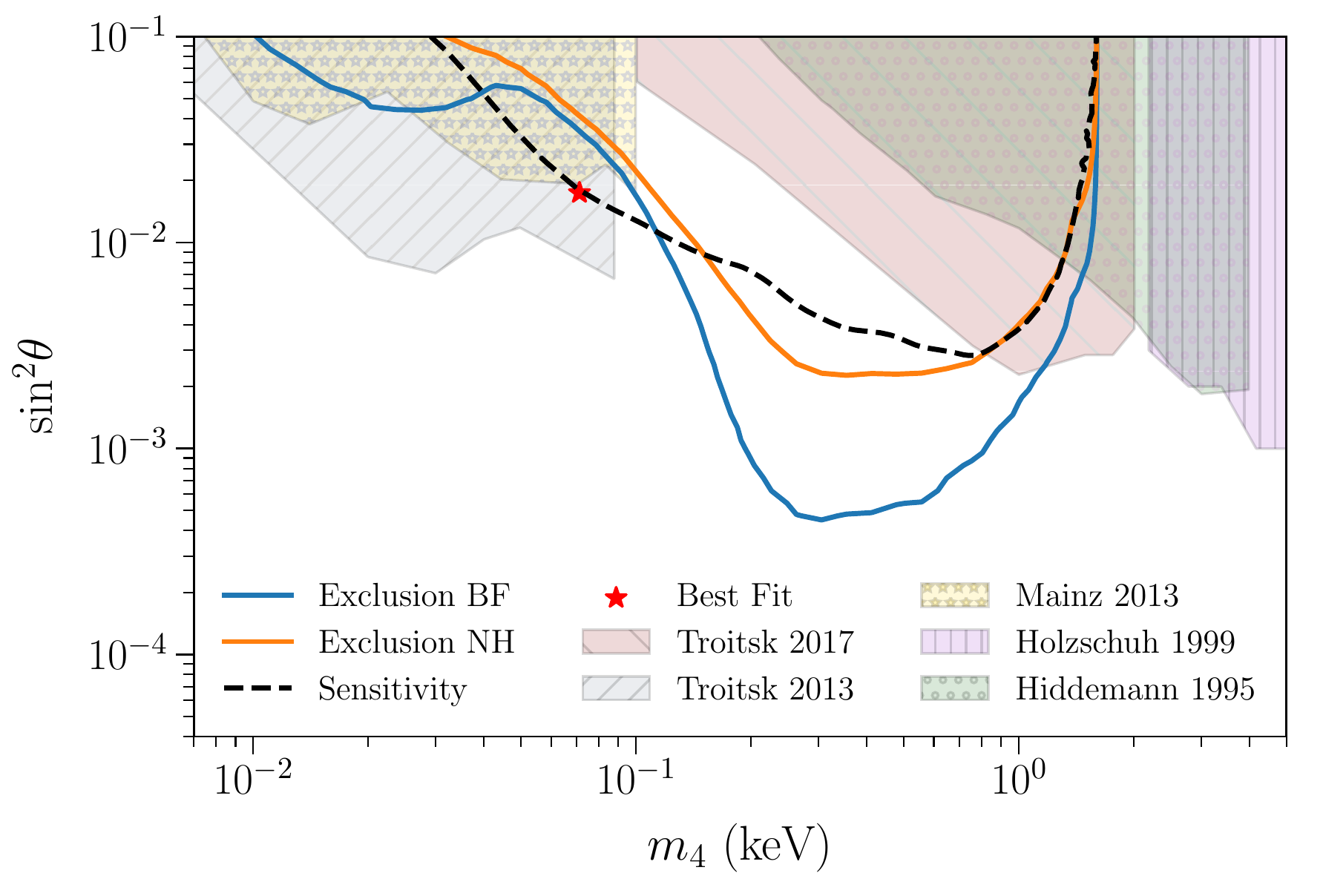}
        \caption{\SI{95}{\percent} C.L. exclusion limit obtained based on the first tritium data set of KATRIN with respect to the best fit (BF) (blue). The exclusion limit obtained by comparing the $\chi^2$ values to the null hypothesis (NH) (orange) is in reasonable agreement with the sensitivity (dashed black). We improve the current laboratory limits~\cite{limit1, limit2, limit3, limit4, limit5} (colored shaded areas) on the active-to-sterile mixing amplitude in a mass range of $0.1\,\mathrm{keV} < m_4 < 1.0\,\mathrm{keV}$ by up to an order of magnitude. As a main result, we quote the exclusion limit with respect to the best fit (blue line).}
        \label{fig:exclusion}
\end{figure}

Following the procedure outlined in section~\ref{sec:exclLimConstr}, we now scan the parameter space ($m_4$, $\sin^2\theta$) and determine the minimal $\chi^2$-value at each grid point. The best fit is found for $m_4 = \SI{71.2}{\electronvolt}$ and $\sin^2\theta = 0.017$. With respect to the null hypothesis, the best fit found at $\Delta \chi^2 = 5.13$, corresponding to a significance of \SI{92.3}{\percent} and a deviation of 2.26 $\sigma$.  Based on this result, we determine the $95\,\%$ C.L. exclusion limit, as shown in figure~\ref{fig:exclusion}. For a mass of $m_4 = \SI{300}{\electronvolt}$ we find the strongest exclusion limit of $\sin^2\theta < 5\cdot10^{-4}$ at $95\,\%$ CL. In addition we display the $95\,\%$ C.L. exclusion limit with respect to the null-hypothesis fit and the $95\,\%$ C.L. exclusion sensitivity.

Finally, we compare our achieved exclusion limit with previous laboratory-based sterile-neutrino searches~\cite{Abdurashitov2017, Radcliffe:1992, PhysRevC.36.1504, PhysRevC.32.2215, OHI1985322, Ohsh93, Muel94, Holz99}. The Troitsk nu-mass experiment provides the leading limit for sterile-neutrino masses of $m_4 < 0.1\,\mathrm{keV}$, based on a re-analysis of their neutrino-mass data~\cite{Belesev:2012hx}. A recent upgrade of the experiment~\cite{Abdurashitov:2015jha} allowed the extension of the measurement interval, setting a new limit for sterile-neutrino masses in the range of $0.1 \, \mathrm{keV} < m_4 < 2\,\mathrm{keV}$~\cite{Abdurashitov2017}. With the analysis presented in this work, we can improve this limit in a mass range of $0.1\,\mathrm{keV} < m_4 < 1.0\,\mathrm{keV}$.

\section{Conclusion and outlook}
In this work we have performed a search for keV-scale sterile neutrinos with a mass of up to  $1.6\,\mathrm{keV}$, based on the first commissioning run of the KATRIN experiment. The analysis includes a careful study of possible systematic uncertainties that occur when extending the nominal KATRIN measurement interval, which is restricted to a region close to the tritium endpoint. 

As a result we exclude an active-sterile mixing amplitude of $\sin^2\theta < 5\cdot10^{-4}$ for a sterile neutrino mass of $m_4 = 300\, \mathrm{eV}$. With this work, we improve currently leading laboratory-based bounds in a mass range of $0.1\,\mathrm{keV} < m_4 < 1.0\,\mathrm{keV}$. This result establishes a major milestone for the keV-scale sterile-neutrino program of KATRIN and sets the groundwork for future high-statistics measurements. 

Currently, a new detector system for KATRIN, the TRISTAN detector, is being developed, which is designed to allow KATRIN to extend the measurement interval to several keV below the endpoint and further improve the laboratory-based sensitivity to keV-scale sterile neutrinos~\cite{Mertens:2018vuu}. This technique will exploit a combination of differential and integral spectral measurements to exclude large classes of systematic effects~\cite{Mertens:2018vuu}.



\section{Acknowledgments}
We acknowledge the support of Helmholtz Association (HGF), Ministry for Education and Research BMBF (05A20PMA, 05A20PX3, 05A20VK3), Helmholtz Alliance for Astroparticle Physics (HAP), the doctoral school KSETA at KIT, and Helmholtz Young Investigator Group (VH-NG-1055), Max Planck Research Group (MaxPlanck@TUM), and Deutsche Forschungsgemeinschaft DFG (Research Training Groups Grants No., GRK 1694 and GRK 2149, Graduate School Grant No. GSC 1085-KSETA, and SFB-1258) in Germany; Ministry of Education, Youth and Sport (CANAM-LM2015056, LTT19005) in the Czech Republic; Ministry of Science and Higher Education of the Russian Federation under contract 075-15-2020-778; and the Department of Energy through grants DE-FG02-97ER41020, DE-FG02-94ER40818, DE-SC0004036, DE-FG02-97ER41033, DE-FG02-97ER41041,  {DE-SC0011091 and DE-SC0019304 and the Federal Prime Agreement DE-AC02-05CH11231} in the United States. This project has received funding from the European Research Council (ERC) under the European Union Horizon 2020 research and innovation programme (grant agreement No. 852845). We thank the computing cluster support at the Institute for Astroparticle Physics at Karlsruhe Institute of Technology, Max Planck Computing and Data Facility (MPCDF), and National Energy Research Scientific Computing Center (NERSC) at Lawrence Berkeley National Laboratory.
\bibliography{literature}

\end{document}